\documentstyle[12pt]{article}

\catcode`\@=11 \@addtoreset{equation}{section}\catcode`\@=12
\newcommand{\ee}{\end{equation}}
\newcommand{\ea}{\end{eqnarray}}

\begin{document}


\bigskip

\centerline{\Large\bf MASS BOUNDS FOR MULTIDIMENSIONAL}
\bigskip
\centerline{\Large\bf CHARGED DILATONIC BLACK HOLES}

\bigskip


\bigskip

\bigskip
\centerline{ \bf U. BLEYER
\footnote{E-mail address: UBleyer@aip.de. This work was supported
by  WIP grant 016659.} and
V. D. IVASHCHUK\footnote{Permanent address:
Center for Surface and Vacuum Research, 8
Kravchenko str., Moscow, 117331, Russia,
e-mail: mel@cvsi.uucp.free.msk.su.
This work was supported in part by DFG grant 436
RUS - 113-7-2.
}
}

\bigskip

\centerline{WIP-Gravitationsprojekt, Universit\"at Potsdam }
\centerline{An der Sternwarte 16 }
\centerline{ D-14482 Potsdam, Germany}

\bigskip

\bigskip

\begin{abstract}
\noindent
The  multidimensional charged dilatonic black hole
solution  with $n$ internal Ricci-flat spaces is considered. The bound
on the mass of the black hole is obtained. In the strong dilatonic
coupling limit the critical mass becomes zero. The case $n =
\infty$ is also considered.
\end{abstract}

\newpage
\section{Introduction}


In [1] Myers and Perry obtained the multidimensional
$O(d+1)$-symmetric analogue of the well-known Reissner-Nordstr\"om
charged black hole solution. In [2,3] the generalization of
the Myers-Perry solution to the case of $n$ Ricci-flat
internal spaces was obtained. Special cases of the solution
[2,3] were considered earlier in the following publications:
[4,5] ($d=2$, $\lambda = 0$), [6] ($d=2$, $n=1$),
[7] ($d \geq 2$, $\lambda = 0$). (In [8] the special case of
solution [2,3] with zero electric and scalar  charges was
obtained.)

The model considered in [2,3] is described by the action
\begin{equation}
S = \int d^{D}x \sqrt{|g|} \{
\frac{1}{2\kappa^{2}} {R}[g] -  \frac{1}{2\kappa^{2}} \partial_{M} \varphi
\partial_{N} \varphi g^{MN} - \frac{1}{4}\exp(2\lambda \varphi) F_{MN}
F^{MN}\},
\end{equation}
where $g = g_{MN} dx^{M} \otimes dx^{N}$ is the metric ,
$F = \frac{1}{2} F_{MN} dx^{M} \wedge dx^{N} = dA $ is the strength
of the electromagnetic field and  $\varphi$ is the scalar field
(dilatonic field).  Here $\lambda$ is the dilatonic coupling constant
and $\kappa$ denotes Einstein's gravitational constant.  The action (1.1)
describes for certain values of the coupling constant $\lambda$
and space-time dimension $D$ a lot of interesting physical models
including standard Kaluza-Klein theory, dimensionally reduced
Einstein-Maxwell theory,  certain sectors of supergravity
theories etc. For
\begin{equation}
\lambda^2 = \frac{1}{2}    , \qquad    D = 10,
\end{equation}
the action  (1.1) describes a part of the bosonic sector for the $N=1$
ten-dimensional Einstein-Yang-Mills  supergravity that occurs in the
low energy limit of superstring theory [9].

In this paper we consider as special case of the solution [2,3]
a charged dilatonic black hole with $n$ internal Ricci-flat
spaces. Objects of such sort are very popular in the literature
(see for example [2,6,10-14]. Here we present
bounds on the mass of this black hole. For $D =4$, $d=2$ such a
bound was obtained by Gibbons and Wells  in [14]. We also
consider an interesting special case of the charged black hole
solution with infinite number of internal spaces $n = \infty$.
We note  that to our knowledge the Einstein
equations in  the infinite-dimensional space were considered
first by Kalitzin [15]. Infinite-dimensional spherically-symmetric
and cosmological solutions were also presented in [3] and [16]
respectively.

\section{Spherically Symmetric Solutions}

The field equations, corresponding to the action (1.1),
have the following form
\begin{eqnarray}
 R_{MN} -\frac{1}{2} g_{MN} R & = & \kappa^{2} T_{MN}, \\
 {\Box} {\varphi} - \frac{\kappa^{2}}{2}\lambda \exp(2\lambda
           \varphi) F_{MN} F^{MN} & = & 0, \\
 \nabla_{M} (\exp(2\lambda \varphi) F^{MN}) & = & 0,
\end{eqnarray}
where
\begin{eqnarray}
T_{MN} = & \frac{1}{\kappa^{2}} (\partial_{M} \varphi \partial_{N}
\varphi  - \frac{1}{2} g_{MN}  \partial_{P} \varphi
\partial^{P} \varphi) \nonumber \\
& + \exp(2\lambda \varphi) (F_{MP} {F_{N}}^{P}
- \frac{1}{4} g_{MN}F_{PQ} F^{PQ}).
\end{eqnarray}

Here we consider the  spherically ${O}(d+1)$-symmetric
solutions of the field equations (2.1)-(2.3) obtained in [2,3].
In this paper the notations of ref. [3] are used.
The solution [4] is defined on the manifold
\begin{equation}
M = M^{(2+d)} \times M_{1} \times \ldots \times M_{n},
\end{equation}
and has the following form
\begin{eqnarray}
g & = &-f_{1}^{(D-3)/{A}(\lambda)} f_{\varphi}^{2\lambda} d\bar{t}
\otimes d\bar{t}
\nonumber \\
&& + f^{-1/{A}(\lambda)}_{1} (f_{2}^{-1}
f_{\varphi}^{2\lambda} f^{2}) ^{1/(1-d)} [f_{2} du \otimes du +
d \Omega^{2}_{d}]   \nonumber  \\
&& + \sum_{i=1}^{n}
f_{1}^{-1/{A}(\lambda)}  \exp(2 A_{i}u +2D_{i}) g^{(i)}, \end{eqnarray}
\begin{equation}
F =  Q f_{1} du \wedge d\bar{t},
\end{equation}
\begin{equation}
\exp \varphi = f_{1}^{(2-D)\lambda/ 2{A}(\lambda)} f_{\varphi}.
\end{equation}
where $M^{(2+d)}$  is a $(2+d)$-dimensional space-time ($d \geq 2$),
the $(M_{i},g^{(i)})$ are Ricci-flat manifolds ($g^{(i)}$ is the metric on
$M_{i}$), $dim M_{i} = N_{i} (i =1, \ldots , n)$,
$d \Omega^{2}_{d}$ is the canonical metric on the $d$-dimensional sphere
$S^{d}$. In (2.6)-(2.8)
\begin{eqnarray}
f_{1} &=& {f_{1}}(u)  = C_{1}(D-2)/ \kappa^{2} Q^{2} {A}(\lambda)
\sinh^{2}(\sqrt{C_{1}} (u-u_{1})) , \\
f_{2} &=& {f_{2}}(u) = C_{2}/(d-1)^{2} \sinh^{2}(\sqrt{C_{2}}(u-u_{2}))
,\\
f_{\varphi} &=& {f_{\varphi}}(u) = \exp (Bu +D_{\varphi}), \\
f &=& {f}(u) = \exp [ \sum_{i=1}^{n} N_{i}( A_{i}u +D_{i})],\\
{A}(\lambda) &=& D-3 + \lambda^{2}(D-2),
\end{eqnarray}
and $Q \neq 0$, $D_{i}$, $D_{\varphi}$, $u_{1}$,
$u_{2}$ are constants and the parameters $C_{1}$, $C_{2}$, $B$, $A_{i}$
satisfy the relation
\begin{eqnarray} \frac{C_{2}d}{d-1} =
&&\frac{C_{1}(D-2)}{D-3 + \lambda^{2}(D-2)} +B^{2} (1+\lambda^{2})
\nonumber   \\
&&+ \frac{1}{d-1} (\lambda B + \sum_{i=1}^{n}
A_{i}N_{i})^{2} + \sum_{i=1}^{n} A_{i}^{2} N_{i}.  \end{eqnarray}

\section{Dilatonic Charged Black Hole}

Now, we are looking for a black hole solution, that means for the
special case of a configuration with external horizon. To do so, we
take  the solution
(2.6)-(2.14)   with the parameters
\begin{equation}
C_{1} = C_{2} = C > 0,  \qquad
u_{2} = 0, \quad u_{1} = - u_{0} < 0,
\end{equation}
\begin{equation}
A_{i}/\sqrt{C} = -1/{A}(\lambda),   \qquad
B/\sqrt{C} = - \lambda  (D-2)/{A}(\lambda).
\end{equation}
Introducing the parameters
\begin{equation}
B_{\pm} = \frac{\kappa |Q| \sqrt{{A}(\lambda)} }{(d-1) \sqrt{D-2}}
\exp(\pm \sqrt{C} u_{0})
\end{equation}
and reparametrizing the time and radial coordinates
\begin{eqnarray}
\bar{t}& =& \frac{\kappa |Q| \sqrt{{A}(\lambda)}\sinh( \sqrt{C} u_{0})
 }{ \sqrt{C(D-2)}} t, \\
r^{d-1}& =& \frac{\kappa |Q| \sqrt{{A}(\lambda)}
\sinh( \sqrt{C} (u + u_{0}))} {(d-1) \sqrt{(D-2)} \sinh( \sqrt{C}
u)},
\end{eqnarray}
we get the following formulas for the solution ($\lambda \neq 0$)
\begin{equation}
g = - f_{+} f_{-}^{1 + 2\alpha_{t}} dt \otimes dt +
 f_{-}^{ 2\alpha_{r}} [\frac{ dr \otimes dr }{ f_{+} f_{-}  }
   + r^{2} d \Omega^{2}_{d}]
 + f_{-} ^{-2/{A}(\lambda)}  \sum_{i=1}^{n}  g^{(i)},
\end{equation}
\begin{equation}
F = Q r^{-d}  dt \wedge dr,
\ee
\begin{equation}
\exp(2\lambda \varphi) = f_{-}^{2\alpha_{t}}.
\ee
Here
\begin{equation}
f_{\pm} = {f_{\pm}}(r) =  1 - \frac{B_{\pm}}{r^{d-1}},
\ee

\begin{equation}
\alpha_{t} = -  \lambda^{2}  (D-2)/{A}(\lambda),  \qquad
\alpha_{r} = \frac{1}{d-1} -  \frac{1}{{A}(\lambda)},
\ee
and the constants $B_{\pm}$ and $Q$ satisfy the relations
\begin{equation}
B_{+} B_{-} = \frac{\kappa^{2} Q^{2}
{A}(\lambda)} {(d-1)^{2}(D-2)}.
\end{equation}
We remind that we consider the case $Q \neq 0$. Due to
eq. (3.3) we have
\begin{equation}
B_{+} > B_{-} > 0.
\end{equation}
In this case the $(2+d)$-dimensional section of the metric (3.6)
has a horizon at
$r^{d-1} = B_{+}$. For $r^{d-1} = B_{-}$  the horizon is absent
($\lambda \neq 0$).
In the limit $\lambda \rightarrow 0$ and  $D \rightarrow
2+d$    we get the Myers-Perry ${O}(d+1)$-symmetric
charged black hole solution [1]. For  $d=2$, $D = 4$  the  solution
coincides (up to redefinitions of the field variables) with the
4-dimensional  dilatonic charged black hole solution [10,11].

\section{Mass Bounds}

The solution (3.6)-(3.11) describes  an ${O}(d+1)$-symmetric charged
dilatonic black hole with a chain of internal Ricci-flat
spaces. The charge of the black hole is $Q$ and the mass $M$ is found
{}from the time component of the metric (3.6) to be
\begin{equation}
2G M =  B_{+} + B_{-}{\beta}(\lambda),
\end{equation}
where
\begin{equation}
{\beta}(\lambda) = 1 + 2 \alpha_{t} = \frac{D-3 - \lambda^{2}(D-2)}
{D-3 + \lambda^{2}(D-2)}
\end{equation}
and
\begin{equation}
G  =  S_{D} \kappa^{2}
\end{equation}
is the effective gravitational constant ($S_{D}$ is defined in [1]).
It is clear that
\begin{equation}
 -1  <  {\beta}(\lambda) < 1
\end{equation}
and $ {\beta}(\lambda) = 0$   for $\lambda_{c}^{2} = (D-3)/(D-2)$.
Using  the relation (4.1) and the inequalities (3.12) and (4.4)
we get
\begin{equation}
 M > M_{c},
\end{equation}
where
\begin{equation}
M_{c} =  \frac{\kappa |Q| (D-3)}{G(d-1) \sqrt{{A}(\lambda)(D-2)}}.
\end{equation}
This can be easily deduced from Fig. 1.


The critical case corresponds to
\begin{equation}
B_{+} = B_{-} = \frac{\kappa |Q| \sqrt{{A}(\lambda)}}
{(d-1) \sqrt{(D-2)}}.
\end{equation}
Formula (4.6) agrees with the corresponding relation from ref. [14] in
the case $D = 4$.

It is not difficult to verify that in the critical case (4.7)
we have a horizon, if and only if
\begin{equation}
\lambda^{2} \leq \frac{1}{d-1} - \frac{D-3}{D-2}.
\end{equation}
For these values of $\lambda$ the inequality (4.5) should be
replaced by
\begin{equation}
 M \geq M_{c}.
\end{equation}

In the strong coupling limit
$\lambda^{2} \rightarrow + \infty$
the critical mass (4.6) tends to zero. A possible interpretation of
this effect seems to be screening by dilatonic field of the
electric charge.

\bigskip

\noindent
{\bf Infinite-dimensional case.}
Now we consider the interesting special case when $n \rightarrow
+ \infty$. In this limit the exact solution of the field equations
taken from (3.6)-(3.11) reads
\begin{equation}
g= - f_{+} f_{-}^{\frac{1 - \lambda^{2}}{1 + \lambda^{2}}}
dt \otimes dt +
 f_{-}^{ \frac{2}{d-1}} \left[\frac{ dr \otimes dr }{ f_{+} f_{-}  }
   + r^{2} d \Omega^{2}_{d} \right]
   +  \sum_{i=1}^{n}  g^{(i)},
\ee
\begin{equation}
F = Q r^{-d}  dt \wedge dr,
\ee
\begin{equation}
\exp(2\lambda \varphi) =  f_{-}^{\frac{1 - \lambda^{2}}{1 + \lambda^{2}}},
\ee
where
\begin{equation}
B_{+} B_{-} = \frac{\kappa^{2} Q^{2}
(1+\lambda^{2})} {(d-1)^{2}}.
\end{equation}
In this case the internal
space scale-factors do not depend on the radial coordinate
but the information about the presence of internal dimensions is
contained in
the $(2+d)$-dimensional part of the metric: this part does not
coincide with the $(D=2+d)$-dimensional solution without
internal spaces. The critical mass is non-zero in this limit:
\begin{equation}
M_{c} =  \frac{\kappa |Q| }{G(d-1) \sqrt{1 +\lambda^{2}}}.
\end{equation}

\section{Conclusion}

We considered the ${O}(d+1)$-symmetric charged dilatonic black hole
solution with $n$ Ricci-flat internal spaces.  We obtained the bound on the
mass of a black hole for all (non-zero) values of the coupling
constant. We found that the critical mass tends to zero in
the strong coupling limit.  We also considered the case of
infinite number of internal spaces : $n = \infty$. In this case
we obtained a non-trivial solution of the field equations and a
non-zero value for the critical mass.

\bigskip

\begin{center} {\bf Acknowledgments}   \end{center}

\noindent
The work was sponsored by the WIP project 016659/p. One of us (V. I.)
was supported by DFG grant 436 RUS 113-7-2 and partly by the Russian
Ministry of Science. V. I. also thanks the colleagues of the WIP
gravitation project group at Potsdam University for their hospitality.

\newpage
\noindent
Figures:

\vspace{3cm}

\noindent
a)

\vspace{9cm}

\noindent
b)

\vfill

\noindent
Fig. 1: Lines of $M = const $ for a) $\beta > 0$ and b) $\beta < 0$

\end{document}